\documentclass[fleqn,11pt,twoside]{article}

\usepackage{amsthm,amsthm,amssymb, color, xcolor,epsfig, graphics, subfigure}

\usepackage{amsmath, graphicx, latexsym, lscape}

\usepackage[pdftex]{hyperref}
\usepackage{tikz-cd}
\usepackage{physics}
\usepackage{float}
\usepackage{listings}
\usepackage{xcolor}
\usepackage{amsfonts}
\usepackage{comment}
\usepackage{mathtools}
\definecolor{codegreen}{rgb}{0,0.6,0}
\definecolor{codegray}{rgb}{0.5,0.5,0.5}
\definecolor{codepurple}{rgb}{0.58,0,0.82}
\definecolor{backcolour}{rgb}{0.95,0.95,0.92}
\lstdefinestyle{mystyle}{
	backgroundcolor=\color{backcolour},   
	commentstyle=\color{codegreen},
	keywordstyle=\color{magenta},
	numberstyle=\tiny\color{codegray},
	stringstyle=\color{codepurple},
	basicstyle=\ttfamily\footnotesize,
	breakatwhitespace=false,         
	breaklines=true,                 
	captionpos=b,                    
	keepspaces=true,                 
	showspaces=false,                
	showstringspaces=false,
	showtabs=false,                  
	tabsize=2
}
\numberwithin{equation}{section}
\usepackage{enumitem}
\usepackage{enumitem,amssymb}
\newlist{todolist}{itemize}{2}
\setlist[todolist]{label=$\square$}
\usepackage{pifont}
\makeatletter
\newcommand{\copyrightnote}[2]{{\renewcommand{\thefootnote}{}
 \footnotetext{\small\it
\begin{flushleft}
 \copyright \ #1   #2  
\end{flushleft}}}}

\newcommand{\Name}[1]{\begin{flushleft}
                       \LARGE \bf #1
                       \end{flushleft}\vspace{-3mm}}

\newcommand{\Author}[1]{\begin{flushleft}
                       \it #1 \end{flushleft}}

\newcommand{\Address}[1]{\begin{flushleft}
                       \it #1 \end{flushleft}}

\newcommand{\Date}[1]{\begin{flushleft}
                      \small  \it #1 \end{flushleft}}

\newcommand{\evenhead}{Author \ name}
\newcommand{\oddhead}{Article \ name}

\renewcommand{\@evenhead}{
\hspace*{-3pt}\raisebox{-15pt}[\headheight][0pt]{\vbox{\hbox to \textwidth
{\thepage \hfil \evenhead}\vskip4pt \hrule}}}
\renewcommand{\@oddhead}{
\hspace*{-3pt}\raisebox{-15pt}[\headheight][0pt]{\vbox{\hbox to \textwidth
{\oddhead \hfil \thepage}\vskip4pt\hrule}}}
\renewcommand{\@evenfoot}{}
\renewcommand{\@oddfoot}{}

\long\def\@makecaption#1#2{%
  \vskip\abovecaptionskip
  \sbox\@tempboxa{\small \textbf{#1.}\ \ #2}%
  \ifdim \wd\@tempboxa >\hsize
    {\small \textbf{#1.}\ \ #2}\par
  \else
    \global \@minipagefalse
    \hb@xt@\hsize{\hfil\box\@tempboxa\hfil}%
  \fi
  \vskip\belowcaptionskip}

\newcommand{\JNMPnumberwithin}[3][\arabic]{%
  \@ifundefined{c@#2}{\@nocounterr{#2}}{%
    \@ifundefined{c@#3}{\@nocnterr{#3}}{%
      \@addtoreset{#2}{#3}%
      \@xp\xdef\csname the#2\endcsname{%
        \@xp\@nx\csname the#3\endcsname .\@nx#1{#2}}}}%
}

\newcommand{\resetfootnoterule} {
  \renewcommand\footnoterule{%
  \kern-3\p@
  \hrule\@width.4\columnwidth
  \kern2.6\p@}
}

\renewcommand{\footnoterule}{}

\makeatother

\theoremstyle{definition}

\newcommand{\be}{\begin{equation}}
	\newcommand{\ee}{\end{equation}}

\def\br{\begin{eqnarray}}
	\def\er{\end{eqnarray}}

\def\({\left(}
\def\){\right)}
\def\[{\left[}
\def\]{\right]}

\def\lie{{\cal G}}
\def\a{\alpha}

\def\cD{{\cal D}}
\def\cE{{\cal E}^{(1)}}

\def\bpsi{\bar{\psi}}

\def\g{\gamma}

\def\l{\lambda}

\def\bnu{\bar{\nu}}
\def\bg{\bar{\gamma}}

\def\O{\Omega}

\def\pa{\partial}

\def\tp0{\Theta_{+}^{(0)}}
\def\tm0{\Theta_{-}^{(0)}}

\def\bp{{\bar \p}}

\def\cS{{\cal S}}

\def\bp{\bar{\psi}}
\def\bchi{\bar{\chi}}

\def\l{\lambda}

\def\bi{\begin{itemize}}
\def\ei{\end{itemize}}

\newcommand{\kdv}{\text{KdV}}
\newcommand{\mkdv}{\text{mKdV}}

\newcommand{\skdv}{\text{SKdV}}
\newcommand{\smkdv}{\text{SmKdV}}

\begin{document}

\renewcommand{\evenhead}{ {\LARGE\textcolor{blue!10!black!40!green}{{\sf \ \ \ ]ocnmp[}}}\strut\hfill 
Y F Adans, A R Aguirre, J F Gomes, G V Lobo and A H Zimerman
}
\renewcommand{\oddhead}{ {\LARGE\textcolor{blue!10!black!40!green}{{\sf ]ocnmp[}}}\ \ \ \ \  
SKdV, SmKdV and supersymmetric
gauge-Miura transformations
}

\thispagestyle{empty}
\newcommand{\FistPageHead}[3]{
\begin{flushleft}
\raisebox{8mm}[0pt][0pt]
{\footnotesize \sf
\parbox{150mm}{{Open Communications in Nonlinear Mathematical Physics}\ \ \ \ {\LARGE\textcolor{blue!10!black!40!green}{]ocnmp[}}
\quad Special Issue 2, 2024\ \  pp
#2\hfill {\sc #3}}}\vspace{-13mm}
\end{flushleft}}

\FistPageHead{1}{\pageref{firstpage}--\pageref{lastpage}}{ \ \ }

\strut\hfill

\strut\hfill

\copyrightnote{The author(s). Distributed under a Creative Commons Attribution 4.0 International License}

\begin{center}

{\bf {\large Proceedings of the OCNMP-2024 Conference:\\ 

\smallskip

Bad Ems, 23-29 June 2024}}
\end{center}

\smallskip

\Name{SKdV, SmKdV flows and their supersymmetric gauge-Miura transformations}

\Author{Y. F. Adans$^{\,1}$, A. R. Aguirre$^{\,2}$, J. F. Gomes$^{\,1}$,  G. V. Lobo$^{\,1}$ and A.H. Zimerman$^{\,1}$}

\Address{$^{1}$ S\~ ao Paulo State University, Institute of Theoretical Physics, IFT-UNESP\\
		Rua Dr. Bento Teobaldo Ferraz 271, 01140-070, São Paulo, SP, Brazil\\[2mm]
$^{2}$ Institute of Physics and Chemistry - IFQ/UNIFEI, Federal University of Itajubá,\\ Av. BPS 1303, 37500-903, Itajubá, MG, Brazil}

\Date{Received March 26, 2024; Accepted April 23, 2024}

\setcounter{equation}{0}

\begin{abstract}

\noindent 
The construction of Integrable Hierarchies  in terms of  zero curvature representation provides a systematic  construction for a series of integrable  non-linear  evolution equations (flows)   which shares a common   affine Lie  algebraic structure.  The integrable hierarchies are   then classified   in terms of a decomposition of the underlying affine  Lie algebra $\hat {\cal {G}} $  into graded subspaces  defined by a grading operator $Q$.
In this paper we shall  discuss  explicitly the simplest case of the affine $\hat {sl}(2)$ Kac-Moody algebra  within the principal gradation  given rise to the KdV and mKdV hierarchies and extend  to  supersymmetric models.

Inspired by the  dressing transformation method,  we have constructed a gauge-Miura  transformation 
mapping mKdV into KdV flows.   Interesting new results   concerns the  negative grade sector of the mKdV hierarchy in which  a double degeneracy of flows (odd and its consecutive even) of mKdV are mapped into a single odd KdV flow.  These results are extended to  supersymmetric  hierarchies  based upon the affine $\hat {sl}(2,1)$ super-algebra.

\end{abstract}

\label{firstpage}


\section{Introduction}

Integrable field theories are very peculiar models  admitting  infinite number of conservation laws and soliton solutions.  The KdV and mKdV  equations are for instance,    typical examples  of such a class of models  and have acted as  prototypes  for many new developments in the subject.  In fact apart from a single equation, say mKdV (or KdV)  a series of other (higher/lower grade) evolution equations of motion  (flows) can be systematically derived  from a zero curvature representation  and a single  universal operator of Lie algebraic origin (Lax operator).  These  evolution equations (flows) form  an {\it Integrable Hierarchy} which  are   constructed   in terms of  a decomposition of the affine Lie algebra $\hat {\lie} $ according to a grading operator denoted by ${Q}$.  The Lax operator is further specified by a second decomposition according to a choice of  a constant grade one  generator $E$.   It therefore  follows  that  an Integrable Hierarchy is  defined by three  Lie algebraic ingredients, namely, i) the affine Lie algebra $\hat {\lie} $, ii) the grading operator $Q$ and iii) the constant  grade one generator $E$ (see for instance \cite{gomes_negative_2009} for a review).

Both  KdV and mKdV  hierarchies are  constructed  from the decomposition of the affine $\hat {\lie} =\hat{sl}(2)$ algebra  according to the  principal grading, $Q$.  Details are explicitly  displayed in the Appendix and,  at this stage it should be pointed out  that  different decomposition generate different  hierarchies. The general structure for a flow ($t_{N},\; N\in \mathbb{Z}$) is given by the zero curvature representation, 
\begin{equation}
	\comm{\pa_x + A_x}{\; \pa_{t_N} + A_{t_N}} = 0
	\label{zcc}
\end{equation}
in terms of a universal Lax operator $L = A_x$.

In section 2, we review the bosonic case. We    first discuss  the KdV and mKdV  positive  grade ($N>0$) sub-hierarchies. It is shown that the graded structure of the decomposition of $\hat {\lie} =\hat{sl}(2)$ algebra implies  $N=2n+1$. Next, the negative sub-hierarchies are proposed and it is shown that for mKdV there are no restrictions upon the  negative integers $N$, i.e., negative odd and negative even sub-hierarchies can be constructed  consistently, \cite{gomes_negative_2009}.  
For the KdV however, only a few equations were known for $N$ negative odd (see \cite{verosky_negative_1991, qiao_negative-order_2012}).  More recently, a general ansatz for constructing  the entire negative odd  KdV sub-hierarchy was  proposed in \cite{adans_negative_2023}. An interesting feature of the mKdV {\it negative even} sub-sector  requires soliton solutions build from  {\it strictly non-zero vacuum}. These are constructed  by gauge transforming  the zero curvature representation (\ref{zcc}) in the vacuum configuration into  a non trivial configuration (dressing method) involving the construction of deformed vertex operators see \cite{gomes_negative_2009}.
Next, the gauge-Miura transformation $S$ mapping $A_{\mu}^ {mKdV} \rightarrow A_{\mu}^ {KdV}$ is  discussed.  The mapping is shown to be one to one as far as $N$ is odd positive. The novelty is a double degeneracy between the two negative  sectors.  It is verified that $S$ maps   the first negative odd and its subsequent  negative flows of the mKdV hierachy into  the  first negative odd of the KdV hierarchy.  The argument generalises to lower graded flows,
\begin{equation}
	\begin{tikzcd}[row sep=0.1em, column sep=3em]
		t_{-2j}^{\mkdv} \arrow[dr, "S"] & \\
		&  t_{-2j+1}^{\kdv} \\ 
		t_{-2j+1}^{\mkdv} \arrow[ur, swap, "S"] 
	\end{tikzcd}
\end{equation}
$j=1, \dots $.Interesting to point out that the flows  $t_{-2j}^{mKdV}$  and $t_{-2j+1}^{mKdV}$  require strictly non-zero and strictly zero vacuum solutions  respectively, while the flow $t_{-2j-1}^{KdV}$ admits both, see \cite{adans_negative_2023}.

An important attempt to  introduce supersymmetry  in integrable models dates  from the pionnering paper by Kuperschmidt \cite{kuper}.  Later  Khovanova \cite{khova}  considered a model based upon $sl(2,1)$ affine algebra and showed  to be invariant under supersymmetry transformation.  Moreover  Miura transformation  connecting the SKdV and SmKdV models was proposed. In this paper we  extend those results to the entire hierarchy  considering both positive and negative  graded sub-hierarchies. 
In section 3 we discuss the construction of the supersymmetric mKdV hierarchy based upon the affine super algebra $\hat {sl}(2,1)$. We first discuss the  positive hierarchy and show that  the flows  are labeled by odd integers and the  equations admit both zero and  non-zero constant  vacuum solutions. Next we consider the negative  sub-hierarchy and show that  there is split into  negative odd and negative even graded  flows.  The negative odd  admits only zero-vacuum whilst the negative even, strictly non-zero vacuum solutions. 

In section 4 we discuss the supersymmetric KdV  hierarchy. It is shown that both SKdV and SmKdV  possess the same algebraic structure  and  the SKdV flows are  constructed  systematically for the positive graded sub-hierarchy.
The negative  sub-hierarchy is then proposed by direct  ansatz  for the time component of two dimensional gauge potential. 

In section 5 we rederive the negative SKdV flows by  gauge-Miura transforming   the potentials from the SmKdV.  The compatibility   
recovers the known Miura transformation  together with   additional  conditions involving  temporal derivatives, temporal Miura transformation.  These  shows  that the gauge-Miura mapping  is one to one as far as positive sub-hierarchy is concerned and  is a two to one for the negative grade sector.

Finally in section 6 we discuss the various implications about such gauge-Miura map in terms of the structure of vacuum solutions.

\section{Review of bosonic case} 

In this section, we shall discuss the construction of the mKdV and KdV hierarchies from an algebraic formalism. Both hierarchies share the same algebraic structure and are related by the well-known Miura transformation.  
A crucial ingredient is the fact that the zero curvature condition  is  preserved under gauge transformation.   We formulate the Miura transformation as gauge transformation acting on zero curvature condition (\ref{zcc}).
We also cover the main results related to the negative flows of both hierarchies and how they are interconnected.

The mKdV and KdV hierarchies can be constructed through gauge potentials that are elements of an affine $\hat {sl(}2)$ algebra endowed with a principal grading operator. The spatial gauge potential for  the mKdV hierarchy is given by
\begin{equation}
	A_x^{\mkdv} = E^{(1)} + A_0
	= \mqty(v & 1 \\ \l & -v)	,
	\label{gauge.x.mkdv}
\end{equation}
where $E^{(1)} = K_1^{(1)}=E_\a^{(0)} + E_{-\a}^ {(1)}  \in \lie_1$ is a grade one constant element, and $A_0 = v \; h_1^{(0)} \in \lie_0$ contains the field of model $v=v(x,t_N)$. The spatial gauge potential for the KdV hierarchy differs from (\ref{gauge.x.mkdv}) due to the algebraic element associated with the field of the theory, namely
\begin{equation}
	A_x^{\kdv} = E^{(1)} + A_{-1}
	= \mqty(0 & 1 \\ \l + J & 0),
	\label{gauge.x.kdv}
\end{equation}
where $A_{-1} = J \; E_{-\a_1}^{(0)} \in \lie_{-1}$ contains the field of model $J=J(x,t_N)$. Given the spatial and temporal gauge potentials, $A_x$ and $A_{t_N}$, respectively, we can derive equations of motion associated with the temporal flow $t_N$ from the \textit{zero curvature condition} (\ref{zcc}) 
where $A_{t_N}$ is constructed from a sum of graded elements of the algebra $\lie$. For the so-called positive sub-hierarchy of mKdV, the temporal gauge potentials are given by
\begin{equation}
	A_{t_N}^{\mkdv} = D^{(N)}_{N} + D^{(N-1)}_{N} + \cdots + D^{(0)}_{N}
	\qquad
	\qquad
	(D^{(a)}_{N} \in \lie_a)
	\label{gauge.t.mkdv.pos}
\end{equation}
whereas for the KdV hierarchy, they are written as follows
\begin{equation}
	A_{t_N}^{\kdv} = \cD^{(N)}_{N} + \cD^{(N-1)}_{N} + \cdots + \cD^{(0)}_{N} + \cD^{(-1)}_{N}
	\qquad
	\qquad
	(\cD^{(a)}_{N} \in \lie_a)
	\label{gauge.t.kdv.pos}
\end{equation}
The various elements $D^{(a)}$ or $\cD^{(a)}$ are recursively determined through the graded decomposition of the zero curvature condition. It is noteworthy that for the highest grade, we obtain the following component for both mKdV and KdV hierarchies
\begin{equation}
	\comm{E^{(1)}}{\cD^{(N)}} = 0
	\qquad
	\text{and}
	\qquad
	\comm{E^{(1)}}{D^{(N)}} = 0,
\end{equation}
which implies that the elements $D^{(N)}$ and $ \cD^{(N)}$ belong to the Kernel $\mathcal{K}_E$ of the element $E^{(1)}$. In particular, the elements of $\mathcal{K}_E$ have odd grades for the $sl(2)$ algebra, i.e., $N = 2m + 1$, with $m \in \mathbb{N}$ (see \eqref{kernel.E1}). Thus, the positive temporal flows of both hierarchies are constrained to  be labeled by odd numbers. 

Concerning the negative temporal flows, $t_{-N}$, we can obtain them through the temporal gauge potentials defining the so-called negative sub-hierarchy of mKdV, given by
\begin{equation}
	A_{t_{-N}}^{\mkdv} = D^{(-N)}_{-N} + D^{(-N+1)}_{-N} + \cdots + D^{(-1)}_{-N}
	\qquad
	\qquad
	(D^{(a)}_{-N} \in \lie_a)
	\label{gauge.t.mkdv.neg}
\end{equation}
similar to the positive temporal flows, we can decompose the zero curvature condition according to its graded structure and determine its various components $D^{(-a)}_{-N}$. 
We therefore obtain  equations of motion which are in general non-local. 
From the  the lowest grade of the decomposition, we find  the  non local equation  for $D^{(-N)}_{-N}$,
\begin{equation}
	\pa_x D^{(-N)}_{-N} + \comm{A_0}{D^{(-N)}_{-N}} = 0 \label{3.37}
\end{equation}
and henceforth find   no restrictions for the negative temporal flows of mKdV hierarchy.

Now, concerning the KdV negative sub-hierarchy, consider the temporal gauge potential given by
\begin{equation}
	A_{t_{-N}}^{\kdv} = \cD^{(-N-2)}_{-N} + \cD^{(-N-1)}_{-N} + \cdots + \cD^{(-1)}_{-N}
	\qquad
	\qquad
	(\cD^{(a)}_{-N} \in \lie_a),
	\label{gauge.t.kdv.neg}
\end{equation}
which can also  be determined by the decomposition of
the zero curvature condition. Unlike the mKdV case  (\ref{3.37}),  we obtain the following equation from the lowest grade,
\begin{equation}
	\comm{A_{-1}}{\cD^{(-N-2)}_{-N}} =0.
\end{equation}
Since $A_{-1} = J(x,t_{-N}) E_{-\a_1}^{(0)}$, the above relation will be only satisfied if the element $\cD^{(-N-2)}_{-N}$ is proportional to $E_{-\a_1}^{(-m)}$. This condition constraints the negative temporal flows  to be $N=2m+1$, i.e., {\it the negative sub-hierarchy of KdV admits only odd flows.}

For the \textbf{mKdV hierarchy}, we have the following equations of motion for the few first temporal flows:

\begin{itemize}
	\item \textbf{for $N=1$:}
	\begin{equation}
		\pa_{t_1} v = \pa_x v,
		\label{mkdv.t1}
	\end{equation}
	which is just the wave equation.
	\item \textbf{for $N=3$:}
	\begin{equation}
		4 \pa_{t_3} v = \pa_x^3 v - 6 v^2 \pa_x v
		\label{mkdv.t3}
	\end{equation}
	we obtain the well-known mKdV equation, which names the whole hierarchy.
	\item \textbf{for $N=-1$:}
	\begin{equation}
		\pa_{t_{-1}} \pa_x \phi = 2 \sinh(2 \phi),
		\label{mkdv.tm1}
	\end{equation}
	we get the well-known sinh-Gordon equation. Here, we have introduced a convenient reparametrization for the field,  $v(x,t_{-N}) = \pa_x \phi(x,t_{-N})$.
	\item \textbf{for $N=-2$:}
	\begin{equation}
		\pa_{t_{-2}} \pa_x \phi = -2 \left(e^{-2 \phi} \pa_x^{-1} e^{2\phi} + e^{2 \phi} \pa_x^{-1} e^{-2\phi} \right),
		\label{mkdv.tm2}		
	\end{equation}
	where the anti-derivative operator is defined by $\pa_x^{-1}f= \int^{x} f(y) d{y}$.
\end{itemize}

For the \textbf{KdV hierarchy}, we have the following equations of motion for the first temporal flows:

\begin{itemize}
	\item \textbf{for $N=1$:}
	\begin{equation}
		\pa_{t_1} J = \pa_x J,
		\label{kdv.t1}
	\end{equation}
	corresponds to the wave equation.
	\item \textbf{for $N=3$:}
	\begin{equation}
		4 \pa_{t_3} J = \pa_x^3 J - 6 J \pa_x J,
		\label{kdv.t3}
	\end{equation}
	gives the well-known KdV equation, which names the hierarchy.
	\item \textbf{for $N=-1$:}
	\begin{equation}
		\pa_{t_{-1}} \pa_x^3 \eta - 4 \pa_x \eta \pa_{t_{-1}} \pa_x \eta - 2 \pa_x^2 \eta \pa_{t_{-1}} \eta = 0,
		\label{kdv.tm1}
	\end{equation}
	where we have defined $J(x,t_{-N}) = \pa_x \eta(x,t_{-N})$. This equation is the counterpart of the sinh-Gordon equation in the KdV hierarchy.
\end{itemize}	

\section{Gauge-Miura transformations} 

The mKdV and KdV equations can be related through the well-known \textit{Miura transformation}, initially proposed in \cite{miura_korteweg-vries_1968}. Other formulations have been proposed to perform this procedure, such as \cite{fordy_factorization_1980, guil_homogeneous_1991}, including mappings { not only between individual equations but also between each flow of both hierarchies.} For the case of $sl(2)$ algebra, this was achieved for the temporal flows in \cite{gomes_miura_2016}, and generalized to  $sl(n+1)$  in \cite{de_carvalho_ferreira_gauge_2021}. More recently, the relation between negative temporal flows has been addressed using an approach in which Miura transformations are performed as gauge transformations that links the gauge potentials of both hierarchies, as discussed in \cite{adans_comments_2023,adans_negative_2023}.

Our proposal to establish a connection between the two  hierarchies involves relating the spatial gauge potentials \eqref{gauge.x.mkdv} and \eqref{gauge.x.kdv} by a gauge transformation, which we have dubbed the \textit{gauge-Miura transformation},
\begin{equation}
	A_x^{\kdv} = S_{\pm} A_x^{\mkdv} S_{\pm}^{-1} + S_{\pm} \pa_x S_{\pm}^{-1},
	\label{gauge.miura.x}
\end{equation}
where
\begin{equation}
	S_{+} = \mqty(1 & 0 \\ v & 0)
	\qquad
	\text{and}
	\qquad
	S_{-} = \mqty(0 & 1 / \l \\ 1 & - v / \l).
	\label{gauge.miura.sl2}
\end{equation}
The spatial gauge potentials \eqref{gauge.x.mkdv} and \eqref{gauge.x.kdv} satisfy the relation \eqref{gauge.miura.x} provided that the mKdV and KdV fields are related as follows,
\begin{equation}
	J = v^2 \mp \pa_x v,
	\label{miura.x.sl2}
\end{equation}
which are precisely  the well-known \textit{Miura transformations}, depending on the respective transformation $S_{\pm}$. These gauge-Miura transformations also act on all temporal potentials of both hierarchies,
\begin{equation}
	A_{t_N}^{\kdv} = S_{\pm} A_{t_N}^{\mkdv} S_{\pm}^{-1} + S_{\pm} \pa_{t_N} S_{\pm}^{-1}.
	\label{gauge.miura.t}
\end{equation}
It was shown in \cite{adans_negative_2023, adans_comments_2023}, that the {\it positive flows} between the mKdV and KdV hierarchies can be related in a {\it one-to-one} correspondence,
\begin{equation} 
	\begin{tikzcd}[row sep=0.2em, column sep=2em]
		t_{N}^{\mkdv} \arrow[r, "S"] & t_{N}^{\kdv} , \quad N=1,3, \dots 
	\end{tikzcd}
	\label{corresp.pos.bos}
\end{equation}
On the other hand, the negative flows satisfy a quite peculiar two-to-one correspondence, namely
\begin{equation}
	\begin{tikzcd}[row sep=0.1em, column sep=3em]
		t_{-N}^{\mkdv} \arrow[dr, "S"] & \\
		&  t_{-N}^{\kdv} \\ 
		t_{-N-1}^{\mkdv} \arrow[ur, swap, "S"] 
	\end{tikzcd}
	\label{corresp.neg.bos}	
\end{equation}
$N=1,3,\dots $. In the case of negative flows an additional relation is required involving time derivatives, \textit{temporal Miura} (see \cite{adans_negative_2023}). For example, eqn. \eqref{gauge.miura.t} for $N=-1$ is valid provided  the following condition is satisfied,
\begin{equation}
	\pa_{t_{-1}}\eta = 2 \; e^{-2\phi(x,t_{-1})}. \label{t-1}
\end{equation}	
On the other hand, if the map occurs between $t_{-2}^{\mkdv}$ and $t_{-1}^{\kdv}$ the relation is completely different
\begin{equation}
	\pa_{t_{-1}}\eta = 4 \; e^{-2 \phi(x,t_{-2})}\pa_x^{-1}(e^{2 \phi(x, t_{-2})}),  \label{t-2}
\end{equation}
and similarly for lower flows.  Notice that (\ref{t-1}) involves solution of mKdV according to flow $t_{-1}$ while  (\ref{t-2}) involves solution according to $t_{-2}$.

These correspondences are particularly useful when analyzing the solutions of these equations. For the positive flows, for each mKdV solution, we can obtain two distinct solutions for the corresponding KdV equation through \eqref{miura.x.sl2}. In contrast, for the negative flows, there is a greater degeneracy of solutions for KdV due to the two-to-one correspondence between the flows. That is, for each pair of negative mKdV flows related to their negative KdV counterpart, there will be four distinct solutions, see \cite{adans_complex_2023}. 

Another relevant fact about both hierarchies is their behavior concerning vacuum solutions, i.e., trivial solutions. For the mKdV hierarchy, the positive flows admit  both zero, $v = 0$ and  non-zero constant vacuum solutions $v = v_0$. 
However, the odd negative flows of the mKdV hierarchies admit only  zero vacuum solutions, while the even negative flows admits strictly  non-zero vacuum solution. 
In \cite{adans_negative_2023}, we discuss that such classification in terms of  vacuum orbits and define two different hierarchies: mKdV-I and mKdV-II.The mKdV-I contains  positive and negative odd flows and is defined in the orbit of a zero vacuum.
The mKdV-II, in turn   contains positive odd and negative even flows and is defined in the orbit of  strictly  non-zero vacuum. 
This analysis can be extended to the KdV hierarchy through Miura transformations as shown in diagrams (\ref{corresp.pos.bos}) and (\ref{corresp.neg.bos})  how the  two different vacuum structures  are couched  within the two, KdV and mKdV  hierarchies.


In the next sections, we will present the main elements for constructing the supersymmetric versions of the mKdV and KdV hierarchies. We derive the equations of motion for the first temporal flows, and show how to derive the supersymmetric version of the gauge-Miura transformations, along with their implications for the correspondence between the flows of both hierarchies.
 
 \section{Super mKdV hierarchy}
 
 In this section we extend  previous  results to  supersymmetric KdV/mKdV hierarchies. The algebraic structure is based on the super Kac-Moody algebra $sl(2,1)$, endowed with the principal grading operator $Q$, and the grade one constant element $\mathcal{E}^{(1)}$, which decompose the algebra into graded subspaces $\lie_n$, of grade $n$. All the details can be found in appendix \ref{algebra.sl(2,1)} (see also \cite{gomes_soliton_2006}, \cite{half-grad}).
 
 Consider the following spatial Lax operator \cite{adans_comments_2023}:
 \begin{equation} \label{axsm}
 	A_x^{\smkdv} = \mathcal{E}^{(1)}+ A_0 + A_{\frac{1}{2}},
 \end{equation}
 where  $\mathcal{E}^{(1)} = K_1^{(1)} + K_2^{(1)} \in \lie_1$ is the grade one constant element, and $A_0 = v \; h_1^{(0)} \in \lie_0$  and  $A_{\frac{1}{2}} = \bar{\psi} \; G_2^{(\frac{1}{2})} \in \lie_{1/2}$ contain the bosonic  and fermionic  fields of the model, $v(x,t_N)$ and $\bar{\psi}(x,t_N)$, respectively. Concerning the  smKdV positive sub-hierarchy, the temporal gauge potential is defined as follows,
 \begin{equation}
 	A_{t_N}^{\smkdv} = D^{(N)}_{N} + D^{(N-1/2)}_{N} + \cdots + D^{(1/2)}_{N} + D^{(0)}_{N}
 	\qquad
 	\qquad
 	(D^{(a)}_{N} \in \lie_a).
 	\label{gauge.t.smkdv.pos}
 \end{equation}
 By decomposing the zero curvature condition \eqref{zcc}, we find the following set of equations 
 \begin{subequations}
 	\begin{align}
 		\comm{\cE}{ D^{(N)}_N} &= 0, \label{highgrade.smkdv}\\
 		\comm{\cE}{D^{(N-\frac{1}{2})}_N} + \comm{A_{\frac{1}{2}}}{D^{(N)}_N} &= 0, \\
 		\comm{\cE}{D^{(N-1)}_N} + \comm{A_{\frac{1}{2}}}{D^{(N-\frac{1}{2})}_N} + \comm{A_0}{D^{(N)}_N} + \pa_x D^{(N)}_N &= 0, \\
 		\nonumber & \; \; \vdots \\
 		\comm{A_{\frac{1}{2}}}{D^{(0)}_N} + \comm{A_0}{D^{(\frac{1}{2})}_N} + \pa_x D^{(\frac{1}{2})}_N - \pa_{t_{N}} A_{\frac{1}{2}} & = 0, \\
 		\comm{A_0}{D^{(0)}_N} + \pa_x D^{(0)}_{N} - \pa_{t_{N}}A_0 &= 0
 	\end{align}
 \end{subequations}
 Since our model is based on a superalgebra, the ansatz for the temporal gauge potential \eqref{gauge.t.smkdv.pos} contains elements with semi-integer grades. From the highest grade \eqref{highgrade.smkdv}, we find that $D^{(N)}_N$ must belong to the kernel of $\mathcal{E}^{(1)}$, implying that $N=2m+1$, with $m \in \mathbb{N}$ (see details in \eqref{kernel.super.E1}). Thus, the positive flows of the super mKdV hierarchy are constrained to be odd. The first non-trivial positive temporal flow occurs when $N=3$, which yields the following equations of motion,
 \begin{subequations}
 	\begin{align}
 		4 \pa_{t_3} v &= \pa_x^3 v - 6 v^2 \pa_x v - 3 \bar{\psi} \pa_x \left( v \pa_x \bar{\psi} \right),
 		\label{smkdv.t3.b}
 		\\[0.2cm]
 		4 \pa_{t_3} \bar{\psi} &= \pa_x^3 \bar{\psi} - 3 v \pa_x \left( v \bar{\psi}\right),
 		\label{smkdv.t3.f}
 	\end{align}	\label{smkdv.t3}
 \end{subequations}
 \noindent
 which are the well-known \textit{Super mKdV equations}. Another relevant flow is for $N=1/2$, which provides us with supersymmetry transformations relating the bosonic and fermionic fields of our model, as follows
 \begin{subequations}
 	\begin{align}
 		\pa_{t_{\frac{1}{2}}} v &= \xi \; \pa_x\bp,
 		\\[0.2cm]
 		\pa_{t_{\frac{1}{2}}} \bp &= \xi \; v,
 	\end{align}	\label{susy.transf.mkdv}
 \end{subequations}
 \noindent where $\xi$ is a fermionic constant. It can be verified that equations \eqref{smkdv.t3.b} and \eqref{smkdv.t3.f} are invariant under these supersymmetric transformations.
 
 To complete the construction of our hierarchy, let us propose the gauge potential for the negative sub-hierarchy of super mKdV as
 \begin{equation}
 	A^{\smkdv}_{t_{-N}} =  D^{(-N)}_{-N} + D^{(-N+\frac{1}{2})}_{-N} + \cdots +D^{(-\frac{1}{2})}_{-N}
 	\qquad
 	\qquad
 	(D^{(a)}_{-N} \in \lie_a).
 	\label{gauge.t.smkdv.neg}
 \end{equation}
 It  can be solved from the zero curvature equation \eqref{zcc} recursively from the following  equations,
 \begin{subequations}
 	\begin{align}
 		\comm{A_0}{ D^{(-N)}_{-N}} + \pa_x  D^{(-N)}_{-N} &= 0, \\
 		\comm{A_{\frac{1}{2}}}{D^{(-N)}_{-N}} + \comm{A_0}{D^{(-N+\frac{1}{2})}_{-N}} + \pa_x D^{(-N+\frac{1}{2})}_{-N} &= 0, \\
 		\nonumber \; \; \vdots & \\
 		\comm{\cE}{D^{(-1)}_{-N}} + \comm{A_0}{D^{(-\frac{1}{2})}_{-N}} - \pa_{t_{-N}} A_0 &= 0, \\
 		\comm{\cE}{D^{(-\frac{1}{2})}_{-N}} - \pa_{t_{-N}} A_{\frac{1}{2}} &= 0.
 	\end{align}
 \end{subequations}

 Unlike the positive flows, in this case we do not get any  restriction upon the values of $N$. The $N=-1$ temporal flow provide us with the following temporal Lax
 \begin{equation} \label{smkdv.tm1.lax}
 	A_{t_{-1}}^{\text{SinhG}} =\cosh{2\phi} \; K_1^{(-1)} + K_2^{(-1)} -  \sinh{2 \phi} \; M_2^{(-1)}  -  \psi\sinh{\phi}  \; F_{2}^{(-\frac{1}{2})} - \psi  \cosh{\phi} \; G_1^{(-\frac{1}{2})}
 \end{equation}
 associated with the super sinh-Gordon equation,
 \begin{subequations}
 	\begin{align}
 		\pa_x \pa_{t_{-1}} \phi &= 2\sinh{2 \phi} - 2\bpsi \psi \sinh{\phi},
 		\\[0.2cm]
 		\pa_{t_{-1}} \bpsi &= 2 \psi \cosh{\phi},
 		\\[0.2cm]
 		\pa_x \psi &= 2 \bp \cosh\phi,
 	\end{align}
 	\label{smkdv.tm1}
 \end{subequations}
 where  $v(x,t_N) = \partial_x \phi(x,t_N)$. For $N=-2$, we obtain,
 \begin{equation} \label{smkdv.tm2.lax}
 	\begin{split}
 		A_{t_{-2}}^{\text{mKdV}} &= M
 		_1^{(-2)}-\frac{e^{\phi} \psi_{-}}{2} \left(F_1^{(-\frac{3}{2})}+G_2^{(-\frac{3}{2})}\right) -\frac{e^{-\phi} \psi_{+}}{2} \left(F_1^{(-\frac{3}{2})}-G_2^{(-\frac{3}{2})}\right) \\&+
 		a_{-}\left(K_1^{(-1)}+M_2^{(-1)}\right)-a_{+}\left(K_1^{(-1)}-M_2^{(-1)}\right)+
 		\left(1 + \psi_{-} \psi_{+} \right)K_2^{(-1)}\\
 		&+ \O_{+} \left(F_2^{(-\frac{1}{2})}+G_1^{(-\frac{1}{2})}\right)+ \O_{-} \left(F_2^{(-\frac{1}{2})}-G_1^{(-\frac{1}{2})}\right)
 	\end{split}
 \end{equation}
 yielding the pair of equations
 \begin{subequations}
 	\begin{align}
 		\pa_{t_{-2}}\pa_x \phi &= -2\left(a_{-}+a_{+}\right)+2 \bp\left(\O_{-}+\O_{+}\right),
 		\\[0.2cm]
 		\pa_{t_{-2}} \bp &= -2 \left(\Omega_{+}-\Omega_{-}\right),
 	\end{align}
 	\label{smkdv.tm2}
 \end{subequations}
 where
 \begin{subequations}
 	\begin{align}
 		\psi_{\pm} &= \pa_x^{-1}\left( e^{\pm \phi} \bp\right),
 		\label{psi.pm}
 		\\[0.2cm]
 		a_{\pm} &= e^{\pm 2\phi} \pa_{x}^{-1} \left[e^{\mp 2\phi} \left( 1+\psi_{\mp} \pa_x \psi_{\pm} \right)\right],
 		\label{a.pm}
 	\end{align}
 	\label{smkdv.tm2}
 \end{subequations}
 and
 \begin{equation} \label{om.pm}
 	\O_{\pm} = \frac{e^{\pm \phi}}{2} \pa_x^{-1} \left[e^{\mp 2\phi} \psi_{\pm}-\bp_{\mp} \mp \pa_x \psi_{\mp} \pa_{x}^{-1} \left(\mp 2a_{\pm}+\bp_{-}\bp_{+}+1\right)\right].
 \end{equation}
 
 In the following section, we will systematize the supersymmetric hierarchy of KdV. 
 
\section{Super KdV Hierarchy}

The super KdV hierarchy shares the same algebraic structure as the super mKdV hierarchy, based on the superalgebra $sl(2,1)$ detailed in Appendix \ref{algebra.sl(2,1)}, and is characterized by the spatial gauge potential,
\begin{equation} 
	A_x^{\skdv} = \mathcal{E}^{(1)}+ A_{-1} + A_{-\frac{1}{2}},
	\label{gauge.x.skdv}
\end{equation}
where $\cE$ is the same given  in eqn. (\ref{axsm}) for the  SmKdV,  $A_{-1} = J \; E_{\a_1}^{(0)} \in \lie_{-1}$ contains the bosonic field,  $J(x,t_N)$ and $A_{-\frac{1}{2}} = \bchi \; ( E_{\a_2}^{(-\frac{1}{2})} - E_{-\a_2}^{(-\frac{1}{2})}) \in \lie_{-1/2}$  the fermionic field  $\bchi(x,t_N)$ of the theory.

The positive sub-hierarchy of sKdV is given by the temporal gauge potential defined as follows,
\begin{equation}
	A_{t_N}^{\skdv} = \cD^{(N)}_{N} + \cD^{(N-1/2)}_{N} + \cD^{(N-1)}_{N} + \cdots + \cD^{(0)}_{N} + \cD^{(-1/2)}_{N} + \cD^{(-1)}_{N} 
	\qquad
	(D^{(n)}_{N} \in \lie_n).
	\label{gauge.t.skdv.pos}
\end{equation}
By decomposing the zero curvature condition \eqref{zcc}, we obtain,
\begin{subequations}
	\begin{align}
		\comm{\cE}{ \cD^{(N)}_N} &= 0, \label{highgrade.skdv}\\
		\comm{\cE}{ \cD^{(N-\frac{1}{2})}_N} &= 0, \\
		\pa_x \cD^{(N)}_N + \comm{\cE}{\cD^{(N-1)}_N} &= 0, \\
		\nonumber & \; \; \vdots \\
		\pa_x \cD^{(-\frac{1}{2})}_N + \comm{A_{-1}}{\cD^{(\frac{1}{2})}_N} + \comm{A_{-\frac{1}{2}}}{\cD^{(0)}_N} - \pa_{t_{N}} A_{-\frac{1}{2}} &=0, \\
		\pa_x \cD^{(-1)}_N + \comm{A_{-1}}{\cD^{(0)}_N} + \comm{A_{-\frac{1}{2}}}{\cD^{(-\frac{1}{2})}_N} - \pa_{t_{N}} A_{-1} &=0.
	\end{align}
\end{subequations}
The highest grade eqn. \eqref{highgrade.skdv} implies  that $\cD^{(N)}_N$ must belong to the kernel of $\mathcal{E}^{(1)}$, and hence  $N=2m+1$, $m \in \mathbb{N}$ (see details in \eqref{kernel.super.E1}). Thus, the positive flows of the super KdV hierarchy are  odd. The first non-trivial positive temporal flow occurs for $N=3$ yielding  the  \textit{Super KdV equation}, 
\begin{subequations}
	\begin{align}
		4 \partial_{t_3} J &= \partial_x^3 J-6 J \partial_x J-3 \bar{\chi} \partial_x^2 \bar{\chi},
		\label{skdv.t3.b}
		\\[0.2cm]
		4 \partial_{t_3} \bar{\chi} &= \partial_x^3 \bar{\chi}-3 \partial_x\left(J \bar{\chi}\right),
		\label{skdv.t3.f}
	\end{align}
	\label{skdv.t3}
\end{subequations}
As before, the positive temporal flow  for $N=1/2$  provides us with supersymmetry transformations  namely,
\begin{subequations}
	\begin{align}
		\pa_{t_{\frac{1}{2}}} J &= \xi \; \pa_x \bchi,
		\\[0.2cm]
		\pa_{t_{\frac{1}{2}}} \bchi &= \xi \; J,
	\end{align}
	\label{susy.transf.kdv}
\end{subequations}
\noindent where $\xi$ is a fermionic constant. The equations of motion \eqref{skdv.t3} remains invariant under supersymmetry transformations \eqref{susy.transf.kdv}. 
Now, to complete the construction of the sKdV hierarchy, let us consider the temporal gauge potential associated to negative grades,
\begin{equation}
	A^{\skdv}_{t_{-N}} =  \cD^{(-N-2)}_{-N} + \cD^{(-N-\frac{3}{2})}_{-N} + \cD^{(-N-1)}_{-N} + \cdots + D^{(-1)}_{-N} + D^{(-\frac{1}{2})}_{-N}
	\qquad
	(D^{(n)}_{-N} \in \lie_n)
	\label{gauge.t.skdv.neg}
\end{equation}
which by the decomposition of \eqref{zcc} leads to the following equations,
\begin{subequations}
	\begin{align}
		\comm{A_{-1}}{\cD^{(-N-2)}_{-N}} &= 0, \label{lowest.grade.skdv.neg} \\
		\comm{A_{-\frac{1}{2}}}{\cD^{(-N-2)}_{-N}} + \comm{A_{-1}}{\cD^{(-N-\frac{3}{2})}_{-N}} &= 0, \\
		\pa_x \cD^{(-N-2)}_{-N} + \comm{A_{-1}}{\cD^{(-N-1)}_{-N}} + \comm{A_{-\frac{1}{2}}}{\cD^{(-N-\frac{3}{2})}_{-N}} &= 0, \\
		\nonumber & \; \; \vdots  \\
		\pa_x \cD^{(-1)}_{-N} + \comm{A_{-\frac{1}{2}}}{\cD^{(-\frac{1}{2})}_{-N}} + \comm{\cE}{\cD^{(-2)}_{-N}} - \pa_{t_{-N}} A_{-1} &= 0, \\
		\pa_x \cD^{(-\frac{1}{2})}_{-N} + \comm{\cE}{\cD^{(-\frac{3}{2})}_{-N}} - \pa_{t_{-N}} A_{-\frac{1}{2}} &= 0, \\
		\comm{\cE}{\cD^{(-1)}_{-N}} &= 0,  \\
		\comm{\cE}{\cD^{(-\frac{1}{2})}_{-N}} &= 0.
	\end{align}
\end{subequations}
From the lowest-grade equation \eqref{lowest.grade.skdv.neg}, we deduce that the $\cD^{(-N-2)}_{-N}$ must be proportional to $E_{-\a_1}^{(m)} \in \lie_{2m+1}$, thus the temporal flows are always odd, i.e., $N=2m+1$. 
Considering  the first negative temporal flow, we find  for  the  temporal gauge potential:
\begin{equation}
	A^{\skdv}_{t_{-1}} = \cD^{(-3)}_{-1} + \cD^{(-\frac{5}{2})}_{-1} + \cD^{(-2)}_{-1} + \cD^{(-\frac{3}{2})}_{-1} + \cD^{(-1)}_{-1}\ + \cD^{(-\frac{1}{2})}_{-1}
	\label{skdv.tm1.lax}
\end{equation}
with
\begin{subequations}
	\begin{align}
		\cD^{{(-3)}}_{-1} &= - \frac{1}{8} \left\{ \pa_x (\pa_{t_{-1}} \pa_x \eta + \pa_x \gamma)  - 2 \pa_x \eta \left( \pa_{t_{-1}}\eta + \gamma \right) \right. 
		\nonumber \\[1mm]
		& \qquad \qquad \qquad \left. + \pa_x \bar{\eta} \left( \bnu_{-} - \bnu_{+} + 2 \bg - \pa_{t_{-1}} \pa_x \bg \right) \right\} \left(K_1^{(-3)} - M_2^{(-3)}\right), \\[3mm]
		\cD^{(-\frac{5}{2})}_{-1} &= \frac{1}{8}\left( \pa_x (\bnu_{+} + \bnu_{-}) - \pa_{t_{-1}} \pa_x^2 \bg + 2 \pa_x \bg \;\g + 2 \pa_x \bg \right) \left( F_2^{(-\frac{5}{2})} + G_1^{(-\frac{5}{2})}\right), \\[3mm]
		\cD^{(-2)}_{-1} &= \frac{1}{4} \left( \pa_{t_{-1}} \pa_x \eta + \pa_x \g \right) \; M_1^{(-2)}, \\[3mm]
		\cD^{(-\frac{3}{2})}_{-1} &= \frac{1}{4} \left( \bnu_{+} - \bnu_{-} - 2 \bg \right) \; F_1^{(-\frac{3}{2})} - \frac{1}{4} \pa_{t_{-1}}\pa_x \bg \; G_2^{(-\frac{3}{2})},
		\\[3mm]
		\cD^{(-1)}_{-1} &= \frac{1}{2} \g \left(K_1^{(-1)} + K_2^{(-1)}\right) + \frac{1}{2} \pa_{t_{-1}} \eta \; K_1^{(-1)} + K_2^{(-1)}, \\[3mm]
		\cD^{(-\frac{1}{2})}_{-1} &= \frac{1}{2} \pa_{t_{-1}} \bg \; F_2^{(-\frac{1}{2})}
	\end{align}
\end{subequations}
Here we have used the following relations,
\begin{subequations}
	\begin{align} \label{skdv.tm1.aux}
		J(x,t_N) &= \pa_x \eta(x,t_N), \quad \quad \quad \bchi(x,t_N) = \pa_x \bg(x,t_N), \quad \quad \quad \pa_x \g = \pa_x \bg \pa_{t_{-1}} \bg,
		\\[2mm]
		\pa_x \bnu_{+} &= \pa_{t_{-1}} \eta \pa_x \bg, \qquad \qquad \quad	\pa_x \bnu_{-} = \pa_x \eta \pa_{t_{-1}} \bg.
	\end{align}
\end{subequations}
yielding  the following pair of equations of motion,
\begin{subequations}
	\begin{align} 
		&\pa_x^3 \pa_{t_{-1}} \eta + \pa_x^2 \g - 2 \pa_x^2 \eta \left( \pa_{t_{-1}} \eta + \g \right) - 4 \pa_x \eta \pa  \pa_x \eta + \pa_x \bg \left( \pa_{t_{-1}} \bg \pa_x \eta - \pa_{t_{-1}} \pa_x \bg - \pa \pa_x^2 \bg \right) \nonumber \\[2mm]
		&+ \pa_x^2 \bg \left( \bnu_{-} - \bnu_{+} + 2 \bg - \pa_{t_{-1}} \pa_x \bg \right) - 2 \pa_x \eta \pa_x \bg \pa_{t_{-1}} \bg = 0,
		\label{skdv.tm1.b}		
		\\[4mm]
		&\pa_x \left(\pa_{t_{-1}} \eta \pa_x \bg + \pa_x\eta \pa_{t_{-1}} \bg - \pa_{t_{-1}} \pa_x^2 \bg + 2 \pa_x \bg \right) + \pa_x \eta \left( \bnu_{+} - \bnu_{-} - 2 \bg + 2 \pa_{t_{-1}} \pa_x \bg  \right) \nonumber \\[2mm]
		&+ \pa_x \bg \left( \pa_{t_{-1}} \pa_x \eta + \pa_x \g \right) = 0.
		\label{skdv.tm1.f}		
	\end{align} \label{skdv.tm1}
\end{subequations}

A natural step now is to determine the connection between the \textit{SmKdV} and \textit{SKdV} hierarchies using a gauge-Miura  transformation.

\section{Gauge Super Miura transformation}

Now that we have established the structure for both 
SmKdV and SKdV Hierarchies, we are able to proceed in determining the Super Miura transformation via gauge transformation, in such a way that it will be possible to map not only the SmKdV equation into SKdV equation, but also the entire hierarchy, including the negative flows. The unifying element employed here  is the mapping the spatial Lax operators  which in the matrix form can be written as
\begin{equation}
	A_x^{\skdv} = \mqty( 
	\sqrt{\l} & 1 & 0 \\
	J + \l & \sqrt{\l} & \bchi \\
	- \bchi & 0 & 2 \sqrt{\l}
	), \qquad A_x^{\smkdv} 
	= \mqty( 
	\sqrt{\l} + v & 1 & - \bpsi \\
	\l & \sqrt{\l} - v & \sqrt{\l} \bpsi \\
	\sqrt{\l} \bpsi & - \bpsi & 2 \sqrt{\l}
	).
\end{equation}

We therefore  search  for  a gauge transformation $\cS$ such that,
\begin{equation}
	A_x^{\skdv} = \cS A_x^{\smkdv} \cS^{-1} + \cS \pa_x \cS^{-1}.
	\label{gauge.smiura.x}
\end{equation}
From the experience gathered  with   the    bosonic case \cite{de_carvalho_ferreira_gauge_2021} we propose the following ansatz,
\begin{equation}
	\cS = \mqty(
	a_{11} & 0 & 0 \\
	0 & a_{22} & 0 \\
	0 & 0 & a_{33}
	)
	+ \mqty(
	0 & 0 & \frac{1}{\sqrt{\l}} a_{13} \\
	0 & 0 & a_{23} \\
	a_{31} & \frac{1}{\sqrt{\l}} a_{32} & 0
	)
	+ \mqty(
	\frac{1}{\sqrt{\l}} b_{11} & \frac{1}{\l} a_{12} & 0 \\
	a_{21} & \frac{1}{\sqrt{\l}} b_{22} & 0 \\
	0 & 0 & \frac{1}{\sqrt{\l}} b_{33}
	)
\end{equation}
which leads to  two different solutions,
\begin{equation}
	\cS_{\pm} = \mqty(
	1 & 0 & 0 \\
	v & 1 & - \bp \\
	\pm \bp & 0 & \pm 1
	).
\end{equation}
These, in turn  leads to the   \textit{Super Miura transformation} relating  SKdV and SmKdV field variables   \footnote{It also possible to obtain a second pair of Super gauge Miura transformation using a second matrix with a different ansatz, given by $	\cS_{2,\pm} = 
	\mqty(
	0 & \frac{1}{\l} & 0 \\
	1 & - \frac{v}{\l} & \bpsi \\
	0 & \mp \bpsi & \frac{\pm 1}{\sqrt{\l}}
	)    $ associated with the two pairs of super Miura transformation: $J^{(2,\pm)} = v^2 + \pa_x v + \bpsi \pa_x \bpsi $ and $	\bar{\chi}^{(2,\pm)}= \mp v \bpsi \mp \pa_x \bpsi$. This choice leads to the Mathieu's Super Miura transformation \cite{mathieu_supersymmetric_1988} under $\bchi \to i \bchi$ and $\bp \to i \bp$ transformation. } \footnote{The fact that exists four different Super Miura transformation is a manifestation of the symmetry of Super mKdV equation under the parity transformation $v \to -v$ and $\bp \to -\bp$.}
\begin{equation} \label{smiura.x.s}
	\begin{split}
		J^{(\pm)} &= v^2 - \pa_x v + \bpsi \pa_x \bpsi ,
		\\
		\bar{\chi}^{(\pm)} &= \mp v \bpsi \pm \pa_x \bpsi.
	\end{split}
\end{equation}
We notice that if we consider the inversion matrix $L_{-}$ acting upon the fermionic subspace, i.e.,	
\begin{equation}
	L_{-} =  \mqty(
	1 &   &   \\
	& 1 &   \\
	&   & - 1
	)
\end{equation}
the following relations holds
\begin{equation*}
	\cS_{-} = L_{-} \; \cS_{+}.
	\qquad
\end{equation*}
Thus, we can proceed assuming $\cS = \cS_{+}$ without loss of generalization. Such fact is very convenient, as is possible to write the gauge-Miura $\cS_{+}$ in the exponential form 
\begin{equation} \label{smiura.exp}
	\cS \equiv	\cS_{+} = e^{ \frac{v}{2} \left(K_1^{(-1)}-M_2^{(-1)}\right) - \frac{\bp}{2} \left(F_2^{(-\frac{1}{2})}+G_1^{(-\frac{1}{2})}\right)  } .
\end{equation}
As expected, we have verified after tedious  but  straightforward calculation that the Super gauge Miura transformation indeed maps	the temporal Lax  $A_{t_{3}}^{\smkdv}$ into $A_{t_{3}}^{\skdv}$  connecting equations \eqref{smkdv.t3}  and \eqref{skdv.t3}. In fact, for higher positive flows it is possible to show using \textit{only} the exponential form \eqref{smiura.exp} together with \eqref{smiura.x.s} that each positive  \textit{SmKdV} flow can be  mapped  into its corresponding  \textit{SKdV} flow:
\begin{equation} 
	\begin{tikzcd}[row sep=0.2em, column sep=2em]
		t_{N}^{\smkdv} \arrow[r, "\cS"] & t_{N}^{\skdv}.
	\end{tikzcd}
	\label{corresp.pos.fer}
\end{equation}

Let us now  extend  our analysis to the negative sector using  the exponential form of $\cS$ (\ref{smiura.exp}). Consider first a generic negative \textit{odd}  flow  $ A_{t_{-2n+1}}^{\smkdv}$ under
the gauge transformation induced by $\cS$.
The  gauge transformation  (\ref{smiura.exp})  yields the following graded structure
\be
\begin{split}
	A_{t_{-2n+1}}^{\skdv} & \equiv \cS \(  D^{(-2n+1)}_{-2n+1}+ D^{(-2n+\frac{3}{2})}_{-2n+1}+\cdots +D^{(-\frac{1}{2})}_{-2n+1}\)\cS^{-1} + \cS \pa_{t_{-2n+1}}\cS^{-1}  \\
	&= {\cD}^{(-2n-1)}_{-2n+1} + {\cD}^{(-2n+\frac{1}{2})}_{-2n+1} +  {\cD}^{(-2n)}_{-2n+1} +\cdots + {\cD}^{(-1)}_{-2n+1} + \cD^{(-\frac{1}{2})}_{-2n+1} \label{kdv1} .
\end{split}
\ee
On the other hand, if we now consider the subsequent negative \textit{even}  flow  $ A_{t_{-2n}}^{\smkdv}$, its lower operator is now proportional to   $D^{(-2n)}_{-2n} \sim M_1^{(-2n)}$,  such that  the final transformation  presents the same algebraic structure, i.e.,
\be
\begin{split}
	\widetilde{A}_{t_{-2n+1}}^{\skdv} &\equiv  \cS \(  D^{(-2n)}_{-2n}+ D^{(-2n+\frac{1}{2})}_{-2n}+\cdots +D^{(-\frac{1}{2})}_{-2n}\)\cS^{-1} + \cS \pa_{t_{-2n}}\cS^{-1}  \\
	&= \widetilde{\cal D}^{(-2n-1)}_{-2n+1} + \widetilde{\cal D}^{(-2n+\frac{1}{2})}_{-2n+1} +  \widetilde{\cal D}^{(-2n)}_{-2n+1} +\cdots + \widetilde{\cal D}^{(-1)}_{-2n+1}+\widetilde{\cal D}^{\left(-\frac{1}{2}\right)}_{-2n+1}.
\end{split}
\label{kdv2}
\ee
Since  the potentials $A_x^{\skdv}$ and $ A_x^{\smkdv} $ are universal within the hierarchies, 
the zero curvature  condition for \eqref{kdv1} and \eqref{kdv2} must yield the same
operator, i.e., 
\be \label{equal_op}
{A}_{t_{-2n+1}}^{\skdv}=\widetilde{A}_{t_{-2n+1}}^{\skdv}  ,
\ee
and the two  gauge potentials provide the  same  evolution equations.
We therefore conclude that, as in the pure bosonic case,   \emph{subsequent negative integer odd and even SmKdV flows collapse into the same negative odd SKdV flow}, which is consistent with the fact that there is no \textit{even} negative flow within the SKdV hierarchy. This  can be  illustrated by
\be\label{corresp.neg.sup}
\begin{tikzcd}[row sep=0.1em, column sep=3em]
	t_{-N}^{\smkdv} \arrow[dr, "\cS"] & \\
	&  t_{-N}^{\skdv} \\ 
	t_{-N-1}^{\smkdv} \arrow[ur, swap, "\cS"] 
\end{tikzcd}
\ee
for $N = 2n - 1$, $n=1,2,\dotsc$.

In order to illustrate such phenomena, we consider an explicit example of  $A_{t_{-1}}^{\smkdv}$ and $A_{t_{-2}}^{\smkdv}$  given by \eqref{smkdv.tm1.lax} and \eqref{smkdv.tm2.lax}.
Indeed the gauge transformation \eqref{gauge.smiura.x} results in the same temporal Lax $A_{t_{-1}}^{\skdv}$ \eqref{skdv.tm1.lax}.


Nevertheless, to obtain such result, it is necessary to introduce an additional information concerning \textit{time derivatives}. If the mapping occurs between $t_{-1}^{\smkdv}$ and  $t_{-1}^{\skdv}$
\begin{equation}
	A_{t_{-1}}^{\skdv} = \cS A_{t_{-1}}^{\smkdv} \cS^{-1} + \cS \pa_x \cS^{-1},
	\label{gauge.smiura.tm1}
\end{equation}
the  \textit{KdV} fields $(\eta,\bg)$ must obey the following relations
\begin{eqnarray}
	\eta_{t_{-1}}&=&2\left(e^{-2\phi}+e^{-\phi} \psi\bp\right) , \label{smiura.tm1.b}\\
	\bg_{t_{-1}}&=&2e^{-\phi} \psi, \label{smiura.tm1.f}
\end{eqnarray}
%
where the mKdV fields in the r.h.s  are  solutions of $t_{-1}$ eqns. (\ref{smkdv.tm1}) .

However, if we are mapping $t_{-2}^{\smkdv}$ into  $t_{-1}^{\skdv}$ 
\begin{equation}
	A_{t_{-1}}^{\skdv} = \cS A_{t_{-2}}^{\smkdv} \cS^{-1} + \cS \pa_x \cS^{-1},
	\label{gauge.smiura.tm2}
\end{equation}
this relation is completely different, namely
\begin{eqnarray}
	\eta_{t_{-1}}&=&4\left(a_{-}+ \O_{-}\bp+\frac{1}{2} \bp_{+} \bp_{-}\right),\label{smiura.tm2.b}\\
	\bg_{t_{-1}}&=&4 \O_{-},\label{smiura.tm2.f}
\end{eqnarray}
where $\bp_{\pm}$, $a_{\pm}$ and $\O_{\pm}$ are given by \eqref{psi.pm}, \eqref{a.pm} and \eqref{om.pm} and satisfy  eqns. (\ref{smkdv.tm2}).

Notice that such set of relations involving KdV variables,  $(\eta,\bg)$ define a distinct set of  \textit{solutions} for equation \eqref{skdv.tm1}. One class of solutions must respect relations \eqref{smiura.x.s} together with the pair \eqref{smiura.tm1.b} and \eqref{smiura.tm1.f}, and the second one must obeys  \eqref{smiura.x.s}, \eqref{smiura.tm1.b} and \eqref{smiura.tm1.f}. This allows us to determine a larger range of solutions for the negative flows within the \textit{SKdV} Hierarchy.
 
 \section{Discussion and further developments} 
 
 In this paper, we have discussed the Miura mapping between the mKdV and KdV flows, extending the approach already used in the pure bosonic case to the supersymmetric case based upon the $sl(2,1)$ affine algebra and the zero curvature representation. The approach employed here involves a gauge transformation acting upon the zero curvature condition. Such a framework has the virtue of relating the entire two hierarchies and henceforth is dubbed the Gauge Super Miura transformation. Using such an approach, we are able to recover well-known results such as the Super Miura transformation \cite{mathieu_supersymmetric_1988}, and also, to discover new ones, such as the coalescence of two subsequent negative flows of SmKdV hierarchy into a single flow of the SKdV hierarchy. We also provide a complete algebraic formulation for the \textit{SKdV} hierarchy and extend the \textit{SmKdV} hierarchy \cite{aguirre_defects_2018} to the negative \textit{even flows.} This result represents a  generalization of the bosonic case proposed in \cite{adans_negative_2023} and demonstrates how an approach focused on algebraic structure together with the formulation of zero curvature enables a general structure that allows the  discover of new results.
 
 On the other hand, it is important to make some comments on such coalescence feature, in particular on the vacuum structure of the equations of motion involved. In order to construct  solutions (or a family of solutions) for an integrable, model such as SmKdV, it is necessary to define a vacuum orbit, i.e., a simple solution which leads to more general ones. In such a scenario, soliton solutions can be obtained by gauge-transforming the gauge potentials in the vacuum, $A_{\mu}^{vac} = A_{\mu}(\phi_0)$, into a nontrivial configuration $A_{\mu}(\phi)$. Such a framework is the basis of the \textit{dressing method} \cite{babelon_affine_1993}. The fact that the flows share the same vacuum orbit is crucial to their involution, and guarantees the existence of the hierarchy \cite{aratyn_integrable_2003}.
 
 It has been shown in several works \cite{gomes_negative_2009, gomes_nonvanishing_2012, adans_twisted_2023} that not only is possible to have a zero vacuum orbit $(v,\bp) = (0,0)$, but it is also possible to use a non-zero vacuum orbit $(v,\bp) = (v_0,\bp_0)$. Now, it turns out that for the SmKdV system, we can verify that equation \eqref{smkdv.t3} admits, besides the zero and non-zero vacuum solutions, intermediary states as $(0,\bp_0)$ and $(v_0,0)$.
 However, this is not true for the negative flows. For instance, in the case of $N=-1$ (Super sinh-Gordon), only a vacuum solution is possible; and for $N=-2$, only non-zero bosonic vacuum is possible, $(v_0,\bp_0)$ with $v_0\neq 0$. In fact, this leads to two different SmKdV hierarchies:
 \begin{itemize} 
 	\item  The SmKdV-I hierarchy has \emph{negative odd} and \emph{positive odd flows}, and  is defined
 	in the orbit of a \textit{bosonic} and \textit{fermionic} \emph{zero vacuum}; 
 	\item The SmKdV-II hierarchy has \emph{negative even} and \emph{positive odd flows}, and is defined
 	in the orbit of a \emph{nonzero bosonic vacuum} and both \emph{zero} or \emph{nonzero fermionic vacuum}.
 \end{itemize}
 For the SKdV Hierarchy, this picture is completely different. As one might notice analyzing equation \eqref{skdv.t3} and \eqref{skdv.tm1}, all the equations shared the same vacuum orbit, either $(J,\bchi) = (0,0)$,  $(J,\bchi) = (J_0,\bchi_0)$, $(J,\bchi) = (0,\bchi_0)$ or $(J,\bchi) = (J_0,0)$, are valid for all the equations\footnote{To verify this for the negative flows of SKdV, it is crucial to use the temporal Super Miura relations \eqref{smiura.tm1.b}, \eqref{smiura.tm1.f} or \eqref{smiura.tm2.b}, \eqref{smiura.tm2.f} due the existence of pure temporal derivatives in the equation of motion}. In such a case, 
 \begin{itemize}
 	\item Each integrable model within the positive and negative part of the SKdV hierarchy admits both zero as well as nonzero  vacuum solutions. 
 \end{itemize}
 This feature certainly explains why the two flows of the negative part of the SmKdV model collapses into one SKdV time flow, since each one posses the vacuum configuration that is necessary for the negative SKdV time flow.
 
 It would be interesting to construct \emph{soliton solutions} for the \textit{SmKdV} and the SKdV hierarchies, by implementing both the \emph{dressing method}, and the Super gauge Miura transformation developed in this work.
 These issues also currently under investigations and will be reported elsewhere.

\subsection*{Acknowledgements}

JFG and AHZ thank CNPq and FAPESP for support. YFA thanks FAPESP for financial support under grant \#2022/13584-0. ARA thanks CAPES for financial support. 

\newpage

\appendix

\section{Algebra $sl(2)$}

\label{algebra.sl(2)}

Consider the $\lie = sl(2)$ centerless Kac-Moody algebra generated by\footnote{
	We employ the following representation for generators:
	\\
	$h_1 = \left( \begin{smallmatrix} 1 & 0 \\ 0 &-1 \\ \end{smallmatrix}\right)$,
	\quad
	$E_{\a_1} = \left( \begin{smallmatrix} 0 & 1 \\ 0 & 0 \\ \end{smallmatrix}\right)$,
	\quad
	$E_{-\a_1} = E_{\a_1}^{\dagger}$.
}
\begin{align}
	\lie = sl(2) = \left\{ h_1^{(m)} = \l^m h_1, \; E_{\a_1}^{(m)} = \l^m E_{\a_1}, \; E_{-\a_1}^{(m)} = \l^m E_{\a_{-1}}\right\}
\end{align}
where $\a_1$ is a simple root. The \textit{principal grading operator} is defined by
\begin{equation}
	Q = 2 \hat{d} + \frac{1}{2} h_1,
	\label{op.grad}
\end{equation}
where $\hat{d}$ is a derivation operator that satisfies
\begin{equation}
	\comm{\hat{d}}{T_a^{(m)}} = m \; T_a^{(m)},
	\quad
	\quad
	\quad
	T_a^{(m)} \in \lie
	\nonumber
\end{equation}
The \textit{principal grading operator} decomposes the algebra in graded subspaces, $\lie  = \bigoplus_i \lie_i$, where
\begin{equation}
	\left[Q, \mathcal{G}_a\right] = a \; \mathcal{G}_a, \qquad \left[\mathcal{G}_a, \mathcal{G}_b\right] \in \mathcal{G}_{a+b},
	\nonumber   
\end{equation}
for $a, b \in  \mathbb{Z}$. For our purposes, the subspaces to consider are:
\begin{equation}
	\begin{split}
		\lie_{2m} &= \left\{ h_1^{(m)} \right\},
		\\
		\lie_{2m+1} &= \left\{ E_{\a_1}^{(m)}, \; E_{-\a_1}^{(m+1)} \right\}.
	\end{split}
	\label{subspace-sl2}
\end{equation}
Another key ingredient to construct our models is a grade one constant element,
\begin{equation}
	\label{E1}
	E^{(1)} = E_{\a_1}^{(0)} + E_{-\a_1}^{(1)}
	= \mqty( 0 & 1 \\
	\l & 0)
\end{equation}
that decomposes the algebra into $\lie = \mathcal{K} \bigoplus \mathcal{M}$ where $\mathcal{K}_E =  \left\{ x \in \lie \, | \, \comm{E}{x} = 0 \right\}$ and $\mathcal{M}_E$ its complement,
\begin{subequations}
	\begin{align}
		\mathcal{K}_E &= \left\{ K_1^{(2m+1)} = E_{\a_1}^{(m)} + E_{-\a_1}^{(m+1)} \right\},
		\label{kernel.E1}
		\\
		\mathcal{M}_E &= \left\{ M_1^{(2m)} = h_1^{(m)}, \; M_2^{(2m+1)} = E_{\a_1}^{(m)} + E_{-\a_1}^{(m+1)}  \right\}.
	\end{align}
\end{subequations}
From \eqref{op.grad} and \eqref{E1}, we can reorganize the graded subspaces \eqref{subspace-sl2} in terms of decomposition kernel-image as follows
\begin{equation}
	\begin{split}
		\lie_{2m} &= \left\{ M_1^{(2m)} \right\},
		\\
		\lie_{2m+1} &= \left\{ K_1^{(2m+1)}, \; M_2^{(2m+1)} \right\}.
	\end{split}
	\label{subs	pace-k+i-sl2}
\end{equation}
The commutation relations of the algebra are given by 
\begin{equation}
	\begin{split}
		\comm{K_1^{(2m+1)}}{K_1^{(2n+1)}} &= 0, \\
		\comm{K_1^{(2m+1)}}{M_1^{(2n)}} &= -2 M_2^{(2m+2n+1)}, \\
		\comm{K_1^{(2m+1)}}{M_2^{(2n+1)}} &= -2 M_1^{(2(m+n+1))},
	\end{split}
	\qquad
	\qquad
	\begin{split}
		\comm{M_1^{(2m)}}{M_1^{(2n)}} &= 0, \\
		\comm{M_1^{(2m)}}{M_2^{(2n+1)}} &= 2 K_1^{(2m+2n+1)}, \\
		\comm{M_2^{(2m+1)}}{M_2^{(2n+1)}} &= 0.
	\end{split}
	\nonumber
\end{equation}

\section{Superalgebra $sl(2,1)$}

\label{algebra.sl(2,1)}

In this section we employ the algebraic formalism to construct an integrable hierarchies with supersymmetry. Consider the $\lie = sl(2,1)$ centerless super Kac-Moody algebra generated by\footnote{
	We employ the following representation for generators:
	\\
	$h_1 = \left( \begin{smallmatrix} 1 & 0 & 0 \\ 0 &-1 & 0 \\ 0 & 0 & 0 \\ \end{smallmatrix}\right)$,
	\quad
	$h_2 = \left( \begin{smallmatrix} 0 & 0 & 0 \\ 0 & 1 & 0 \\ 0 & 0 & 1 \\ \end{smallmatrix}\right)$,
	\quad
	$E_{\a_1} = \left( \begin{smallmatrix} 0 & 1 & 0 \\ 0 & 0 & 0 \\ 0 & 0 & 0 \\ \end{smallmatrix}\right)$,
	\quad
	$E_{\a_2} = \left( \begin{smallmatrix} 0 & 0 & 0 \\ 0 & 0 & 1 \\ 0 & 0 & 0 \\ \end{smallmatrix}\right)$,
	\\
	\quad
	$E_{\a_1+\a_2} = \left( \begin{smallmatrix} 0 & 0 & 1 \\ 0 & 0 & 0 \\ 0 & 0 & 0 \\ \end{smallmatrix}\right)$,
	\quad
	$E_{-\a_1} = E_{\a_1}^{\dagger}$,
	\quad
	$E_{-\a_2} = E_{\a_2}^{\dagger}$,
	\quad
	$E_{-(\a_1+\a_2)} = E_{(\a_1+\a_2)}^{\dagger}$.
}
\begin{align}
	L_0 &= \left\{ h_1^{(m)} = \l^{m} h_1, \; h_2^{(m)} = \l^{m} h_2, \; E_{\pm \a_1}^{(m)} = \l^{m} E_{\pm \a_1} \right\},
	\nonumber
	\\ 
	\nonumber
	\\
	L_1 &= \left\{ E_{\pm \a_2}^{(m)} = \l^m E_{\pm \a_2}, \; E_{\pm (\a_1+\a_2)}^{(m)} = \l^m E_{\pm (\a_1 + \a_2)} \right\},
	\nonumber
\end{align}
where $\l \in \mathbb{C}$ and $m \in \mathbb{N}$. The $L_0$ and $L_1$ are called the bosonic and fermionic parts of algebra, respectively, that satisfying the following relations
\begin{equation}
	\comm{L_0}{L_0} \subset L_0,
	\qquad
	\comm{L_0}{L_1} \subset L_1,
	\qquad
	\comm{L_1}{L_1} \subset L_0.
	\nonumber
\end{equation}
The \textit{principal grading operator} is defined by
\begin{equation}
	Q = 2 \hat{d} + \frac{1}{2} h_1
	\label{op.grad_super}
\end{equation}
where $\hat{d}$ is a derivation operator that satisfies
\begin{equation}
	\comm{\hat{d}}{T_a^{(m)}} = m \; T_a^{(m)},
	\quad
	\quad
	\quad
	T_a^{(m)} \in \lie
	\nonumber
\end{equation}
The \textit{principal grading operator} decomposes the algebra in graded subspaces, $\lie  = \bigoplus_i \lie_i$, where
\begin{equation}
	\left[Q, \mathcal{G}_a\right] = a \; \mathcal{G}_a, \qquad \left[\mathcal{G}_a, \mathcal{G}_b\right] \in \mathcal{G}_{a+b},
	\nonumber   
\end{equation}
for $a, b \in  \mathbb{Z}$. For our purposes, the subspaces to consider are:
\begin{equation}
	\begin{split}
		\lie_{2m} &= \left\{ h_1^{(m)} \right\},
		\\
		\lie_{2m+\frac{1}{2}} &= \left\{ E_{\a_2}^{(m+1/2)}, \; E_{\a_1 + \a_2}^{(m)}, \; E_{-\a_2}^{(m)}, \; E_{-\a_1 - \a_2}^{(m+1/2)} \right\},
		\\
		\lie_{2m+1} &= \left\{ E_{\a_1}^{(m)}, \; E_{-\a_1}^{(m+1)}, \; h_2^{(m+1/2)} \right\},
		\\
		\lie_{2m+\frac{3}{2}} &= \left\{ E_{\a_2}^{(m+1/2)}, \; E_{\a_1 + \a_2}^{(m+1/2)}, \; E_{-\a_2}^{(m+1/2)}, \; E_{-\a_1 - \a_2}^{(m+1)} \right\}.
	\end{split}
	\label{subspace}
\end{equation}
Another key ingredient to construct our models is a grade one constant element,
\begin{equation}
	\label{super_E1}
	\mathcal{E}^{(1)} = E_{\a_1}^{(0)} + E_{-\a_1}^{(1)} + h_1^{(1/2)} + 2 h_2^{(1/2)}
	= \mqty( \sqrt{\l} & 1 & 0 \\
	\l & \sqrt{\l} & 0 \\
	0 & 0 & 2 \sqrt{\l}),
\end{equation}
that decomposes the algebra into $\lie = \mathcal{K} \bigoplus \mathcal{M}$ where $\mathcal{K}_{\mathcal{E}} =  \left\{ x \in \lie \, | \, \comm{E}{x} = 0 \right\}$ and $\mathcal{M}_{\mathcal{E}}$ its complement,
\begin{equation}
	\begin{split}
		\mathcal{K}_\text{Bose} = \mathcal{K}_{\mathcal{E}} \cap L_0 &= \left\{ K_1^{(2m+1}, \; K_2^{(2m+1)} \right\},
		\\
		\mathcal{K}_\text{Fermi} = \mathcal{K}_{\mathcal{E}} \cap L_1 &= \left\{ F_1^{(2m+1/2)}, \; F_2^{(2m+3/2)}  \right\},
	\end{split}
	\quad
	\begin{split}
		\mathcal{M}_\text{Bose} = \mathcal{M}_{\mathcal{E}} \cap L_0 &= \left\{ M_1^{(2m)} \right\},
		\\
		\mathcal{M}_\text{Fermi} = \mathcal{M}_{\mathcal{E}} \cap L_1 &= \left\{ G_1^{(2m+3/2)}, \; G_2^{(2m+1/2)} \right\},
	\end{split}
	\label{kernel.super.E1}
\end{equation}
where the bosonic generators are be defined as
\begin{equation}
	\begin{split}
		K_1^{(2m+1)} &= E_{\a_1}^{(m)} + E_{-\a_1}^{(m+1)},\\
		K_2^{(2m+1)} &= h_1^{(m+1/2)} + 2 h_2^{(m+1/2)},
	\end{split}
	\qquad
	\begin{split}
		M_1^{(2m)} &=  h_1^{(m)},\\
		M_2^{(2m+1)} &= E_{\a_1}^{(m)} - E_{-\a_1}^{(m+1)},   
	\end{split}
	\nonumber
\end{equation}
and the fermionic generators are
\begin{subequations}
	\begin{align*}
		F_{1}^{(2 m+\frac{1}{2})}&=\left(E_{\alpha_2}^{(m+\frac{1}{2})}+E_{\alpha_1+\alpha_2}^{(m)}\right)+\left(E_{-\alpha_2}^{(m)}+E_{-\left(\alpha_1+\alpha_2\right)}^{(m+\frac{1}{2})}\right),
		\\
		F_{2}^{(2 m+\frac{3}{2})}&=\left(E_{\alpha_2}^{(m+1)}+E_{\alpha_1+\alpha_2}^{(m+\frac{1}{2})}\right)-\left(E_{-\alpha_2}^{(m+\frac{1}{2})}+E_{-\left(\alpha_1+\alpha_2\right)}^{(m+1)}\right),
		\\
		G_{1}^{(2 m+\frac{3}{2})}&=\left(E_{\alpha_2}^{(m+1)}-E_{\alpha_1+\alpha_2}^{(m+\frac{1}{2})}\right)+\left(E_{-\alpha_2}^{(m+\frac{1}{2})}-E_{-\left(\alpha_1+\alpha_2\right)}^{(m+1)}\right),
		\\
		G_{2}^{(2 m+\frac{1}{2})}&=\left(E_{\alpha_2}^{(m+\frac{1}{2})}-E_{\alpha_1+\alpha_2}^{(m)}\right)-\left(E_{-\alpha_2}^{(m)}-E_{-\left(\alpha_1+\alpha_2\right)}^{(m+\frac{1}{2})}\right).
	\end{align*}
\end{subequations}
From \eqref{op.grad} and \eqref{super_E1}, we can reorganize the graded subspaces \eqref{subspace} in terms of decomposition kernel-image as follows
\begin{equation}
	\begin{split}
		\lie_{2m} &= \left\{ M_1^{(2m)} \right\},
		\\
		\lie_{2m+\frac{1}{2}} &= \left\{F_1^{(2m+\frac{1}{2})}, \; G_2^{(2m+\frac{1}{2})} \right\},
		\\
		\lie_{2m+1} &= \left\{ K_1^{(2m+1)}, \; K_2^{(2m+1)}, \; M_2^{(2m+1)} \right\},
		\\
		\lie_{2m+\frac{3}{2}} &= \left\{ F_2^{(2m+\frac{3}{2})}, \; G_1^{(2m+\frac{3}{2})}\right\}.
	\end{split}
	\label{subspace-k+i}
\end{equation}
The commutation relations  of the algebra are then given by
\begin{equation}
	\begin{split}
		\comm{K_1^{(2m+1)}}{K_1^{(2n+1)}} &= 0, \\
		\comm{K_1^{(2m+1)}}{K_2^{(2n+1)}} &= 0, \\
		\comm{K_1^{(2m+1)}}{M_1^{(2n)}} &= -2 M_2^{(2m+2n+1)}, \\
		\comm{K_1^{(2m+1)}}{M_2^{(2n+1)}} &= -2 M_1^{(2(m+n+1))}, \\
		\comm{K_2^{(2m+1)}}{K_2^{(2n+1)}} &= 0, \\
		\comm{K_1^{(2m+1)}}{F_1^{(2n+1/2)}} &= F_2^{(2(m+n)+3/2)}, \\
		\comm{K_1^{(2m+1)}}{F_2^{(2n+3/2)}} &= F_1^{(2(m+n+1)+1/2)}, \\
		\comm{K_1^{(2m+1)}}{G_1^{(2n+3/2)}} &= - G_2^{(2(m+n+1)+1/2)} , \\
		\comm{K_1^{(2m+1)}}{G_2^{(2n+1/2)}} &= - G_1^{(2(m+n)+3/2)}, \\
		\comm{K_2^{(2m+1)}}{F_1^{(2n+1/2)}} &= - F_2^{(2(m+n)+3/2)}, \\
		\comm{K_2^{(2m+1)}}{F_2^{(2n+3/2)}} &= - F_1^{(2(m+n+1)+1/2)}, \\
		\comm{K_2^{(2m+1)}}{G_1^{(2n+3/2)}} &= - G_2^{(2(m+n+1)+1/2)}, \\
		\comm{K_2^{(2m+1)}}{G_2^{(2n+1/2)}} &= - G_1^{(2(m+n)+3/2)}, 
	\end{split}
	\qquad
	\qquad
	\begin{split}
		\comm{K_2^{(2m+1)}}{M_1^{(2n)}} &= 0, \\
		\comm{K_2^{(2m+1)}}{M_2^{(2n+1)}} &= 0, \\
		\comm{M_1^{(2m)}}{M_1^{(2n)}} &= 0, \\
		\comm{M_1^{(2m)}}{M_2^{(2n+1)}} &= 2 K_1^{(2m+2n+1)}, \\
		\comm{M_2^{(2m+1)}}{M_2^{(2n+1)}} &= 0.\\
		\comm{M_1^{(2m)}}{F_1^{(2n+1/2)}} &= - G_2^{(2(m+2)+1/2)}, \\
		\comm{M_1^{(2m)}}{F_2^{(2n+3/2)}} &= - G_1^{(2(m+2)+3/2)}, \\
		\comm{M_1^{(2m)}}{G_1^{(2n+3/2)}} &= - F_2^{(2(m+2)+3/2)}, \\
		\comm{M_1^{(2m+1)}}{G_2^{(2n+1/2)}} &= - F_1^{(2(m+2)+1/2)}, \\
		\comm{M_2^{(2m+1)}}{F_1^{(2n+1/2)}} &= - G_1^{(2(m+n)+3/2)}, \\
		\comm{M_2^{(2m+1)}}{F_2^{(2n+3/2)}} &= - G_2^{(2(m+n+1)+1/2)}, \\
		\comm{M_2^{(2m+1)}}{G_1^{(2n+3/2)}} &= F_1^{(2(m+n+1)+1/2)}, \\
		\comm{M_2^{(2m+1)}}{G_2^{(2n+1/2)}} &= F_2^{(2(m+n)+3/2)}.
	\end{split}\nonumber
\end{equation}
and the anti-commutations relations, given by 
\begin{equation}
	\begin{split}
		\acomm{F_1^{(2m+1/2)}}{F_1^{(2n+1/2)}} &= 2 (K_1^{(2m+2n+1)} + K_2^{(2m+2n+1)}), \\
		\acomm{F_1^{(2m+1/2)}}{F_2^{(2n+3/2)}} &= 0, \\
		\acomm{F_1^{(2m+1/2)}}{G_1^{(2n+3/2)}} &=  -2 M_1^{(2(m+n+1))}, \\
		\acomm{F_1^{(2m+1/2)}}{G_2^{(2n+1/2)}} &= -2 M_2^{(2m+2n+1)}, \\ 
		\acomm{F_2^{(2m+3/2)}}{F_2^{(2n+3/2)}} &= - (K_1^{(2(m+n+1)+1)} + K_2^{(2(m+n+1)+1)}), \\
		\acomm{F_2^{(2m+3/2)}}{G_1^{(2n+3/2)}} &= 2 M_2^{(2(m+n+1)+1)}, \\
		\acomm{F_2^{(2m+3/2)}}{G_2^{(2n+1/2)}} &= 2 M_1^{(2(m+n+1))}, \\ 
		\acomm{G_1^{(2m+3/2)}}{G_1^{(2n+3/2)}} &= -2 (K_1^{(2(m+n+1)+1)} - K_2^{(2(m+n+1)+1)}), \\
		\acomm{G_1^{(2m+3/2)}}{G_2^{(2n+1/2)}} &= 0, \\ 
		\acomm{G_2^{(2m+1/2)}}{G_2^{(2n+1/2)}} &= 2 (K_1^{(2m+2n+1)} + K_2^{(2m+2n+1)}).
	\end{split}
	\nonumber
\end{equation}

\label{lastpage}
\end{document}